\documentclass[12pt,epsf]{article}

\usepackage[dvips]{graphicx}

\newcommand{\beq}{\begin{equation}}
\newcommand{\eeq}{\end{equation}}
\newcommand{\bea}{\begin{eqnarray}}
\newcommand{\eea}{\end{eqnarray}}

\begin{document}
\thispagestyle{empty}
\vspace*{-15mm}
{\bf OCHA-PP-327}\\

\vspace{15mm}
\begin{center}
{\Large\bf
Hydrodynamics on non-commutative space \\
--A step toward hydrodynamics of granular materials--
}
\vspace{7mm}

\baselineskip 18pt
{\bf Mayumi Saitou$^{1}$, Kazuharu Bamba$^{1, 2}$ and Akio Sugamoto$^{1}$}
\vspace{2mm}

{\it
$^1$Department of Physics, Graduate School of Humanities and Sciences, \\
   Ochanomizu University, Tokyo 112-8610, Japan}\\
{\it
$^2$Leading Graduate School Promotion Center, \\
Ochanomizu University, Tokyo 112-8610, Japan}\\

\vspace{10mm}
\end{center}
\begin{center}
\begin{minipage}{14cm}
\baselineskip 16pt
\noindent
\begin{abstract}
  Hydrodynamics on non-commutative space is studied based on a formulation of hydrodynamics by Y. Nambu in terms of Poisson and Nambu brackets.  Replacing these brackets by Moyal brackets with a parameter $\theta$, a new hydrodynamics on non-commutative space is derived. It may be a step toward to find the hydrodynamics of granular materials whose minimum volume is given by $\theta$.  To clarify this minimum volume, path integral quantization and uncertainty relation of Nambu dynamics are examined.   
\end{abstract}


\end{minipage}
\end{center}

\baselineskip 18pt
\def\thefootnote{\fnsymbol{footnote}}
\setcounter{footnote}{0}


\section{Introduction}
In 1973, Y.~Nambu proposed a generalized Hamiltonian dynamics, in which the  usual phase space spanned by a canonical pair $(p, q)$ is generalized to that spanned by more than three canonical variables $(x_1, x_2, \dots, x_n)$~\cite{Nambu dynamics}.  The simplest generalization is a three dimensional phase space of $(x_1, x_2, x_3)$, where Hamilton's equation of motion is written in terms of two Hamiltonians, $H_1$ and $H_2$, as follows:
\bea
\frac{dx_i}{dt} = \frac{\partial(x_i, H_1, H_2)}{\partial(x_1, x_2, x_3)} \quad (i=1, \cdots, 3).
\eea
For the time development of an observable $O(x_1, x_2, x_3)$, we have
\bea
\frac{dO}{dt}= \frac{\partial(O, H_1, H_2 )}{\partial (x_1, x_2, x_3)}.
\eea
The right-hand sides are written in terms of Jacobians.  In the usual Hamilton dynamics, Liouville theorem states that the phase space volume $dp \wedge dq$ occupied by an ensemble of dynamical systems is preserved in time. 
The generalization of this to the $n$-dimensional phase space is easy.
Therefore, in the generalized ($n$-dimensional) Hamiltonian dynamics, being called Nambu dynamics now, the phase space volume $dx_1 \wedge dx_2 \dots \wedge dx_n$ occupied by an ensemble of systems is temporarily preserved.  The dynamics incorporate naturally the infinite dimensional local symmetries of the volume preserving diffeomorphisms whose transformations $(x_1, \dots, x_n) \to (x'_1, \dots, x'_n)$ preserves the Jacobian,
\bea
\frac{\partial(x'_1, x'_2, \dots, x'_n)}{\partial (x_1, x_2, \dots, x_n)}=1. 
\eea 
For the two-dimensional phase space case, $\partial(A, B)/\partial (q, p)$ is the Poisson bracket, and for the case of phase space having more than three canonical variables, we call $\partial(A_1, A_2, \dots, A_n)/\partial (x_1, x_2, \dots, x_n)$ Nambu bracket.

The quantization of this generalized Hamiltonian dynamics, or the quantization of the Nambu bracket, was tried in the paper of 1973~\cite{Nambu dynamics}.  Since then many people have tried to quantize the Nambu brackets by using various methods~\cite{Quantization of ND 1, Hoppe thesis, S-S-DFR, Takhtajan, Dito:1996xr, Curtright:2002fd, ALMY-K, Kawamura:2003cw, Quantization of ND 3, Ho:2013iia}. 

In the background of Nambu dynamics, there exists a volume preserving diffeomorphisms for an ensemble of dynamical systems, so that it naturally fits to the incompressible fluid dynamics, where an ensemble of ingredients of fluid moves in time, keeping its occupying volume.  Therefore, it is quite natural that recently Nambu reformulated hydrodynamics in terms of Poisson brackets in two spacial dimensions and Nambu brackets in three spacial 
dimensions~\cite{Nambu}.  He considered of course an incompressible fluid. 

In this paper, we investigate a hydrodynamics on non-commutative space based on the formulation of hydrodynamics by Nambu.  We construct a new hydrodynamics on non-commutative space through the replacement of the Poisson and Nambu brackets by the Moyal ones.  This is a method invented by Moyal~\cite{Moyal} about the quantization, so that we use it to quantize the space or to find the quantum Nambu brackets.  Since we have to clarify the meaning of the Moyal bracket, we discuss a relationship between the Moyal product and the path integral quantization of a toy model.  In the toy model the Moyal product may reproduces the expectation value of the quantum theory.  

Our final aim is to produce the hydrodynamics describing the motion of 
granular materials whose minimum volume is expressed by a model parameter $\theta$ in the Moyal bracket.  The physics of granular materials is an interesting topic and is now rapidly developing~\cite{granular materials}.  
To clarify the minimum volume, we examine the quantization of the Nambu dynamics in the path integral formulation.  In three dimensional phase space, the quantum Nambu dynamics is a closed string theory.  In this way the uncertainty relation which gives the basis of minimum volume, is clarified. 
We note that 
the extension of Lagrangian formulation of non-commutative perfect fluids 
has been explored in Ref.~\cite{Prof-Vancea}, 
and diffusion in non-commutative geometries has been studied 
in Ref.~\cite{Prof-Pedraza}. 
In addition, uncertainty relations in non-commutative space-time~\cite{Curtright:2001jn} and an application of hydrodynamics like one by Nambu for D-branes~\cite{Curtright:2003je} have been investigated. 

It is true that different ways of quantization give different hydrodynamics.  So, it is interesting to consider different hydrodynamics on different non-commutative spaces with different quantization methods, and compare the obtained results to the experimental data which seems to be compiled so far for various granular materials.  This is, however, beyond the scope of this paper.
The organization of this paper is as follows. 
In Section 2, we review the hydrodynamics by Nambu. 
In Section 3, we formulate a new hydrodynamics on non-commutative space,
starting from the hydrodynamics by Nambu. 
In Section 4, we compare the Moyal product with the expectation value in the path integral quantization of a toy model.
In Section 5, we examine the path integral quantization of Nambu dynamics in general and clarify its uncertainty relation.
Our investigations are finally concluded in Section 6. 

\section{Nambu's hydrodynamics}


The continuity equation of fluid is given in terms of density $\rho(x;t)$ and velocity $\mbox{\boldmath $v$}(x;t)$ of the fluid by
\bea
\dot \rho(x;t)+ \mbox{\boldmath $\nabla$} (\rho(x;t) \mbox{\boldmath $v$}(x;t))=0,
\eea
which becomes in the incompressible case ($\rho=$ const) as
\bea
\mbox{\boldmath $\nabla$} \mbox{\boldmath $v$}(x;t)=0. \label{2}
\eea
Here, the dot denotes the time derivative of $\partial/\partial t$, 
and $\nabla$ is the differential operator as $\nabla \equiv (\partial/\partial x_1, \partial/\partial x_2, \partial/\partial x_3)$. 
Then, we can introduce stream functions, one function $\varphi (x_1,x_2;t)$ in two spacial dimensions and two functions $\varphi_{1} (x_1,x_2,x_3;t)$ and $\varphi_{2} (x_1,x_2,x_3;t)$ in three spacial dimensions, and express velocity fields so as to satisfy the continuity equation (\ref{2}) as follows:
\bea
v_i=\dot x_i &=& \{x_i,\varphi \}_P \quad \mbox {(i=1, 2 for 2D)} ,\\
v_i=\dot x_i &=& \{x_i,\varphi_1, \varphi_2 \}_N \quad \mbox {(i=1, 2, 3 for 3D)} ,
\eea
where Poisson bracket and Nambu bracket are defined by Jacobian, 
\bea
\hspace{-5mm}
\{A_1, A_2, \dots, A_n\}&=&\frac{\partial (A_1, A_2, \dots, A_n)}{\partial (x_1, x_2, \dots, x_n)} \nonumber \\
\hspace{-5mm}
&=&\sum_{i_1, i_2, \dots, i_n=1}^{n} \epsilon^{i_1, i_2, \dots, i_n} \partial_{i_1} A_1(x;t)\partial_{i_2} A_2(x;t) \dots \partial_{i_n} A_n(x;t),
\eea
where $\epsilon^{i_1, i_2, \dots, i_n} $ is the Levi-Civita tensor or the totally anti-symmetric tensor.
The case of $n=2$ is Poisson bracket and that of $n=3$ is Nambu bracket. 

Nambu considered that the position of an element of fluid $x_i(t)$ $(i=1, \cdots,  n) $ at time $t$ is parameterized by its initial (material) coordinates $(\sigma_1, \sigma_2, \dots, \sigma_n)$ at $t=0$, that is,
\bea
x_i(t)=x_i(\sigma_1, \sigma_2, \dots, \sigma_n; t) \quad (i=1, \cdots, n).
\eea
Then, the incompressibility condition is given by
\bea
\frac{\partial (x_1, x_2, \dots, x_n)}{\partial (\sigma_1, \sigma_2, \dots, \sigma_n)}=1. \label{Jacobian=1}
\eea
Full usage of this condition he derived the Navier-Stokes equation, where the Jacobian in terms of $(\sigma_i (i=1, \cdots, n))$ which appears in the beginning is replaced finally by the Jacobian in terms of $(x_i (i=1, \cdots, n))$, Poisson and Nambu brackets, due to (\ref{Jacobian=1}).  

The equations of motion of two dimensional (2D) incompressible fluid ($i=1, 2$) so derived by Nambu are
  \bea
\rho \left( \{x_i, \dot \varphi\}+\{ \{x_i, \varphi\}, \varphi\} \right)+\epsilon^{ij}\{p, x_j\} - \eta \Delta \{x_i, \varphi\}=0,
\eea
while in three dimensional (3D) fluid ($i=1, 2, 3$) they read
\bea
&&
\rho \left(\{x_i, \dot \varphi_1, \varphi_2\}+\{x_i, \varphi_1, \dot \varphi_2\}+\{\{x_i, \varphi_1, \varphi_2\}, \varphi_1, \varphi_2\} \right)
\nonumber \\
&&
{}+\frac{1}{2} \epsilon^{ijk} \{p, x_j, x_k\} - \eta \Delta \{x_i, \varphi_1, \varphi_2\}=0, \label{Nambu equation}
\eea
where $p$ is the pressure, but the external potential $V$ may be included into $p$ like $p+V$, 
$\Delta$ is the Laplacian, 
and the index of shear viscosity $\eta$ is introduced. 
These equations are identical to the usual Navier-Stokes equations,
\bea
\rho \frac{D \mbox{\boldmath $v$}}{Dt} + \mbox{\boldmath $\nabla$} p - \eta \Delta \mbox{\boldmath $v$}=0,  \label{NS}
\eea 
where the Lagrangian derivative is
\bea
\frac{D \mbox{\boldmath $v$} }{Dt} &=& \frac{\partial\mbox{\boldmath $v$}}{\partial t}+ (\mbox{\boldmath $v$} \cdot \mbox{\boldmath $\nabla$}) \mbox{\boldmath $v$} \nonumber \\
&=&\frac{\partial\mbox{\boldmath $v$}}{\partial t}+\mbox{\boldmath $\nabla$}(\frac{1}{2}\mbox{\boldmath $v$}^2)+\mbox{\boldmath $\omega$} \times \mbox{\boldmath $v$},
\eea
and $\mbox{\boldmath $\omega$} =\mbox{\boldmath $\nabla$} \times \mbox{\boldmath $v$}$ is the vorticity.  In two dimensions we have to choose $\mbox{\boldmath $\omega$}=(0, 0, \omega)$ as usual.

It is instructive to derive the Nambu equations (\ref{Nambu equation}) explicitly, starting from the Navier-Stokes equations (\ref{NS}).

\section{Hydrodynamics on non-commutative space}

Now, we introduce the Moyal product and the Moyal bracket and are going to replace Poisson and Nambu brackets by the Moyal brackets.  

Moyal product or $\ast$-product is defined with a constant parameter $\theta_{ab}$ by~\cite{Moyal}
\bea
A(x)\ast B(x)= \left. \exp \left( \frac{i}{2!} \theta_{ab} \frac{\partial^2}{\partial y^a\partial z^b}\right) A(y)B(z) \right|_{y, z \to x}, 
\eea
and its natural generalization to the three $\ast$-product with a parameter $\theta_{abc}$ is
\bea
A(x)\ast B(x)\ast C(x)= \left. \exp \left( \frac{i}{3!} \theta_{abc} \frac{\partial^3}{\partial y^a\partial z^b\partial u^c}\right) A(y)B(z)C(u) \right|_{y, z, u \to x}. 
\eea
By taking simply $\theta_{ab}=\epsilon_{ab} \theta_2$, and $\theta_{abc}=\epsilon_{abc} \theta_3$, then what we have introduced is a parameter with the dimension of area for $\theta_2$, or volume for $\theta_3$.

The Moyal bracket is defined as follows:
\bea
[A(x), B(x)]_{M}=\sum_{A, B} \epsilon_{AB} ~A(x) \ast B(x), \label{Moyal2}
\eea
and 
\bea
[A(x), B(x), C(x)]_{M}=\sum_{A, B, C} \epsilon_{ABC} ~A(x) \ast B(x) \ast C(x). \label{Moyal3}
\eea

Now we are going to replace the Poisson bracket in two dimensional hydrodynamics and the Nambu bracket in the three dimensional hydrodynamics by the corresponding Moyal brackets as follows:
\bea
\{A, B\}_P &\rightarrow& \frac{1}{i \theta_2} [A, B]_M, \\
\{A, B, C\}_N &\rightarrow& \frac{1}{i \theta_3} [A, B, C]_M.
\eea
Then, we will arrive at a new hydrodynamics having a parameter $\theta_2$ or $\theta_3$ which may be related to the size of the granular materials consisting of the fluid.  

The result of the replacement: all the single Moyal brackets are identical to the Poisson bracket or the Nambu bracket, and the difference arises only in the double Moyal brackets, that is, for the two dimensional hydrodynamics, 
\bea
& & [ [x_i, \varphi (x)]_{M}, \varphi (x)]_{M} \nonumber \\
&=& \left. \{ \{x_i, \varphi(x)\}, \varphi(x)\} -\frac{(\theta_2)^2}{24}\left( \left( \partial_{y_1}\partial_{z_2}-\partial_{y_2}\partial_{z_1}\right)^3 v_i(y) \varphi(z) \right) \right|_{y, z \to x} 
\nonumber \\
&+& O\left((\theta_2)^4\right),
\eea
and in the three dimensional hydrodynamics, the difference appears in
\bea
& & 
[ [x_i, \varphi_1 (x), \varphi_2 (x)]_{M}, \varphi_1 (x), \varphi_2 (x)]_{M} \nonumber \\
&=&\{ \{x_i, \varphi_1(x), \varphi_2 (x)\}, \varphi_1(x), \varphi_2 (x)\} \nonumber \\
&-& \left. 
\frac{(\theta_3)^2}{3!} \epsilon_{v_i, \varphi_1, \varphi_2} \left( \left( \sum_{abc} \frac{\partial^3}{\partial_{y_a}\partial_{z_b}\partial_{u_c} }\right)^3 v_i (y)\varphi_1(z)\varphi_2(u)\right) \right|_{y, z, u \to x} 
\nonumber \\
&+& O\left((\theta_3)^4\right).
\eea

Now the Navier-Stokes equations of motion in the non-commutative space with $O(\theta^2)$ corrections are given by
\bea
\rho \frac{D \mbox{\boldmath $v$}}{Dt} + \mbox{\boldmath $\nabla$} p - \eta \Delta \mbox{\boldmath $v$}=K,  \label{NC-NS}
\eea
where $O(\theta^2)$ correction $K$ reads
\bea
\hspace{-5mm}
K&=& \left. \frac{(\theta_2)^2}{24} \rho \left( \partial_{y_1}\partial_{z_2}-\partial_{y_2}\partial_{z_1}\right)^2 \sum_{a=1, 2} \partial_{y_a} \mbox{\boldmath $v$} (y) \mbox{\boldmath $v$}_a (z) \right|_{y, z \to x} 
\quad 
\mbox{(2D)}, \\
\hspace{-5mm}
K&=& \left. \frac{(\theta_3)^2}{3!} \rho \, \epsilon_{v, \varphi_1, \varphi_2} \left( \left( \frac{1}{3!}\sum_{abc} \frac{\partial^3}{\partial_{y_a}\partial_{z_b}\partial_{u_c} }\right)^3  \mbox{\boldmath $v$}(y) \varphi_1(z)\varphi_2(u)\right) \right|_{y, z, u \to x}
\quad 
\mbox{(3D)}.
\eea 
The velocity in 3D is related to stream functions $\varphi_1$ and $\varphi_2$ 
as
\bea
\mbox{\boldmath $v$}^a= \frac{1}{2} \epsilon_{abc}\frac{\partial(\varphi_1, \varphi_2)}{\partial (x_b, x_c)}.
\eea

\section{Moyal product and path integral of a toy model}

We have to understand the uncertainty relation, or the possibility of introducing by $\theta$ a minimum size to the element of the fluid.
In case of two dimensions, the meaning of the Moyal product is clear. We know that the quantum mechanical operator algebra exists behind.
Introduce two operators $\hat A (\hat x)$ and $\hat B (\hat x)$, and assume the operator relation for the variables
\bea
[\hat x_a, \hat y_b]=i \theta \delta_{ab} \quad (a, b=1,\ 2) .
\eea
Here we put the hat on operators, and the commutator is the usual one in the operator algebra.
Assuming the following Fourier expansion
\bea
\hat A (\hat x)&= &\int \frac{dp}{(2\pi)^2} e^{-i p \hat x} A(p), \\
\hat B (\hat y)&=& \int \frac{dq}{(2\pi)^2} e^{-i q \hat y} B(q),
\eea
which fixes the operator ordering of $\hat x$ in $\hat A (\hat x)$ and $\hat B (\hat y)$.  Then, we can prove that
\bea
\hat A (\hat x)\hat B (\hat x)= \left. A(x) \ast B(x) \right|_{x \to \hat x}. 
\label{operator and M product}
\eea
Therefore, the Moyal bracket is faithfully represent the commutation relation 
of the operator algebra, or
\bea
\left. 
[A(x), B(x)]_M \right|_{x \to \hat x} 
= [\hat A (\hat x), \hat B (\hat x)] .
\eea
Now we can understand the uncertainty relation which is valid also in the hydrodynamics of the non-commutative space, 
\bea
\langle(\Delta x)^2 \rangle^{1/2} \langle (\Delta y)^2 \rangle^{1/2} \geq \theta_2/2. \label{uncertainty in 2D}
\eea
Then, we may consider that each element of the fluid to have a minimum area $\theta/2$, or the fluid to consist of a granular material.

Next we compare the Moyal product and the expectation value in the path integral quantization of a toy model.
The expectation value $\langle O \rangle$ in terms of the path integral method of a toy model is given by
\bea
\langle O(x) \rangle \propto \int D X DY O(X, Y) \exp \left( \frac{1}{\theta_2} \left[iXY- \frac{1}{2}(X^2 + Y^2) \right] \right) .
\eea
If we consider X is a momentum and Y is a coordinate, this simplified model may represent the quantum mechanics, while if we consider both X and Y are coordinates, it may represent the non-commutative space.  Here $\frac{1}{2}(X^2 + Y^2)$ is a toy Hamiltonian.  
Notice that even after Wick rotation the phase factor remains as a phase factor.   
The phase factor
\bea
\exp \left( \frac{i}{\hbar} \int p dq \right)
\eea
is the origin of quantum algebra, so that a phase factor 
\bea
\exp \left( \frac{i}{\theta_2} XY \right)
\eea
in the Moyal product is the origin of non-commutativity in space.  
The expectation value can be calculated perturbatively as
\bea
&&
\langle O(X, Y) \rangle = O\left(\frac{1}{i}\frac{\partial}{\partial J_X}, \frac{1}{i}\frac{\partial}{\partial J_Y}\right) 
\nonumber \\
&& \left. 
\hspace{25mm}
\times 
\exp \left( \frac{i}{\theta_2} \left[\frac{1}{i}\frac{\partial}{\partial J_X}\frac{1}{i}\frac{\partial}{\partial J_Y}- \frac{\theta_2}{2}(J_X^2 + J_Y^2) \right] \right) \right|_{J_X, J_Y \to 0}. 
\eea
This shows that $X$ and $Y$ in the operator $O$ is contracted with $X$ and $Y$ in the phase factor with the propagator  $\langle XX \rangle = \langle YY \rangle = \theta_2$, so that we may understand that
\bea
\langle O(X, Y) \rangle=O(X, Y)_{\ast}= \exp \left( i \theta_2 \frac{\partial^2}{\partial X\partial Y}\right) O(X, Y).
\eea

Here we have to comment on a relation between the ordering of factors in the Moyal product and the time ordering of them in the path integral.
Consider the product $A(X_{+}) \ast B(X_{-})$, then, this corresponds to the time ordering in the path integral, or the path integral over $A(X_{+}=X(t_{+})) B(X_{-}=X(t_{-}))$ with  $t_{+}>t_{-}$.  Finally we have to take the limit $t_{+}, t_{-} \to t$.  The phase factor in this case is more precisely 
\bea
\exp \left(-\frac{i}{2! \theta_2}\epsilon_{ab} \left(X^a dX^b \right) \right)=
\exp \left( \frac{i}{2! \theta_2} \epsilon_{ab} \left(X^a_{+} X^b_{-} - (a \leftrightarrow  b) \right) \right).
\eea
Therefore, the Moyal product is understood to be equal to the path integral expectation value of the toy model.  In general the Moyal product and the quantum expectation value may differ, because of other interactions than the mass terms or the Gaussian damping factors.

Now we go to 3D hydrodynamics.  How the uncertainty relation appears in this case is an interesting issue, but the discussion of it is postponed to the next section where the quantization of the Nambu dynamics will be discussed.  Here, we simply compare the results of Moyal product and the path integral, using a toy model.  We consider
\bea
&&
\langle O(X, Y, Z) \rangle \propto \int D X DY DZ O(X, Y, Z) 
\nonumber \\
&&
\hspace{30mm}
\times 
\exp \left(\left[\frac{i}{\theta_3} XYZ- \frac{1}{2(\theta_3)^{2/3}} (X^2 + Y^2+ Z^2) \right] \right).
\eea
This includes the three dimensional phase space factor.
The propagator in this case is $(\theta_3)^{2/3}$, so that we have 
\bea
\langle O(X, Y, Z) \rangle= O(X, Y, Z)_{\ast}=\exp \left( i \theta_3 \frac{\partial^3}{\partial X\partial Y\partial Z} \right) O(X, Y, Z).
\eea
About the ordering of the Moyal product, we have to examine the phase factor more explicitly,
\bea
\exp \left(-\frac{i}{3!\theta_3} \epsilon_{abc} \int X^a \frac{\partial(X^b, X^c)}{\partial (\sigma, t)} d\sigma dt \right).
\eea
If we restrict to an infinitesimal rectangular region formed by four corners $(A, B, C, D)$ the coordinates of which are
\bea
\left[
\begin{array}{cc}
D(\sigma, t),  &  A(\sigma, t-\Delta t)  \\
C(\sigma-\Delta \sigma, t),  & B(\sigma-\Delta \sigma, t-\Delta t)
\end{array}
\right],
\eea
then the phase factor becomes
\bea
&&
\exp \left( \frac{i}{3!\theta_3} \epsilon_{abc} \left( X^a(B)X^b(A)X^c(D) + X^a(D)X^b(C)X^c(B) 
\right. \right. 
\nonumber \\ 
&&
\left. \left. 
\hspace{60.8mm}
{}
- X^a(C)X^b(B)X^c(A) \right) \right).  \label{3D ordering}
\eea
In the next section we will understand that the quantum theory in 3D is a closed string theory.  In this terminology, a closed string $C$ develops in time by a deformation in which a portion $\overrightarrow{BA}$ of a closed string $C$ is replaced by $\overrightarrow{BCDA}$ by a rectangular deformation $\delta C=\overrightarrow{ABCDA}$.  The time evolution is done in this way, so that the ``area" of the rectangular $\overrightarrow{ABCDA}$ plays the role of ``time".  Accordingly, the concept of the time ordering in 2D should be changed in 3D.  The ordering in 3D is the path ordering associated with the infinitesimal closed path $\delta C$, the boundary curve of the rectangular $\overrightarrow{ABCDA}$.  If we take the limit 
$\Delta t \Delta \sigma \to 0$, the phase factor becomes
\bea
\exp \left(\frac{i}{3!\theta_3} \epsilon_{abc} P \left( X^aX^bX^c\right) \right),
\eea
where $P$ denotes the path ordering with respect to the closed path $\delta C$, or the boundary curve of the rectangular $\overrightarrow{ABCDA}$.  Now, the ordering of the Moyal product $A(X) \ast B(Y) \ast C(Z)$ means the path ordering of the three operators $(X, Y, Z)$ in this sense.
So, the Moyal product may give the expectation values in the path integral of the toy mode also in 3D, but it may not reproduce all of the quantum properties in more general cases, because of the possible existence of additional interactions. However, the Moyal product reproduces the essential part of the quantum, or the non-commutative properties.

\section{Path integral Quantization of Nambu dynamics and its uncertainty relation}
Action of Nambu dynamics is given by Takhtajan in theorem 7 of \cite{Takhtajan}, but this action was already known by Nambu in the Hamilton-Jacobi formulation of the string theory \cite{Nambu H-J}. 
The action is 
\bea
S_n=\int X_1  dX_2 \wedge \dots \wedge dX_n- H_1 dH_2 \wedge \cdots \wedge dH_{n-1} \wedge dt,
\eea
where $t$ is time.  The fact that the minimum configuration of the action gives the equation of motion of Nambu dynamics is shown by \cite{Takhtajan}.  Let us study the case of $n=3$.  
\bea
S_3=\int X dY \wedge dZ- H_1 dH_2 \wedge dt.
\eea
As was pointed in \cite{Takhtajan} and \cite{Nambu H-J}, this is not a point particle theory, but a closed string theory the configuration of which is specified by a circle (2-cycle) $C(\sigma, t)$ on the two dimensional plane $(Y, Z)=(X_2, X_3)$, namely
\bea
C(\sigma, t)=\{(Y(\sigma, t), Z(\sigma, t)\}  ~~\mbox{with}~~  (0 \le \sigma \le 2\pi, -\infty \le t \le +\infty), 
\eea
where the closed string means $C(0, t)=C(2\pi, t)$.  
Now, the path integral quantization of $n=3$ Nambu dynamics is given by the following partition function  
\bea
Z \propto \int DX(\sigma, t) DY(\sigma, t) DZ(\sigma, t) \exp \left(\frac{i}{\theta_3} S_3 [X(\sigma, t), Y(\sigma, t), Z(\sigma, t)] \right).
\eea
Notice that this is the path integral in phase space $(X, Y, Z)$, and is not in configuration space.  But, if the momentum $X$ is integrated out, then the usual path integral expression in configuration space is obtained.  A path is specified by a configuration, $\{X(\sigma, t), C(\sigma, t)\}=\{X(\sigma, t), Y(\sigma, t), Z(\sigma, t)\}$ parameterized by two parameters, $\sigma$ and $t$.

Now we introduce the wave functional $\Psi [C(\sigma); t]$.  Here we consider $\Psi$ to depend on the coordinates $Y$ and $Z$, but not on the momentum $X$.  This is correct usually, since due to the uncertainty relation which will appear shortly, we are not able to specify all of these (X, Y, Z) certainly at a given time $t$.
Then, $\Psi_{\alpha, \beta} [C(\sigma); t]$ is given by
\bea
& &\Psi_{\alpha, \beta} [C(\sigma); t]  \nonumber \\
&\propto&\int_{C_{\alpha, \beta} (\sigma), t_0}^{C(\sigma), t}  DX(\sigma, t) DY(\sigma, t) DZ(\sigma, t) \exp \left(\frac{i}{\theta_3} \left[ \int X dY \wedge dZ- H_1 dH_2 \wedge dt \right]\right)  \nonumber \\
&\times& \Psi [C_{\alpha, \beta}(\sigma); t_0] ,
\eea
where $C_{\alpha, \beta}$ denote the initial configurations (shapes) of the closed strings at $t_0$.  The wave functional $\Psi_{\alpha}$ depends on the initial configurations which may label the state vectors $\vert \Psi_{\alpha} [C(\sigma); t] \rangle$.

The amplitude of an observable $O(X(\sigma), C(\sigma); t)$ is given by
\bea
&&
\langle \alpha \mid \hat O \mid \beta \rangle \propto \int DX(\sigma, t) DY(\sigma, t) DZ(\sigma, t)
\nonumber \\
&& \hspace{30mm}
{}\times 
\Psi_\alpha [C(\sigma); t]^{\dagger}  O(X(\sigma), C(\sigma); t)\Psi_\beta [C(\sigma); t]. \label{amplitude}
\eea
Following Feynman \cite{Feynman}, we can read off the operator algebra from the path integral expression.  We introduce the area $A(C)$ of the circle,
\bea
A(C)= \oint_C Y \wedge dZ, 
\eea
and the functional derivative $\delta / \delta C(\sigma) $ corresponding to the path deformation at $\sigma$, $\delta C(\sigma)$, appeared in the last section. 
It is usually defined as 
\bea
\frac{\delta}{\delta C(\sigma)}= \lim_{\delta C(\sigma) \to 0} \frac{\Psi [C(\sigma)+\delta C(\sigma)]-\Psi [C(\sigma)]}{\mbox{area of}~  \delta C(\sigma)}.
\eea
We understand  
\bea
\frac{\delta A(C)}{\delta C(\sigma)} =1, 
\eea
so we have
\bea
\frac{\delta}{\delta C(\sigma)} \Psi [C(\sigma); t]&=& \frac{i}{\theta_3} X(\sigma, t) \Psi [C(\sigma); t], \\
\frac{\partial}{\partial t} \Psi [C(\sigma); t]&=& -\frac{i}{\theta_3}\left( \oint_C H_1dH_2 \right) \Psi [C(\sigma); t].
\eea
If we choose $O(X, Y, Z)$ in Eq.~(\ref{amplitude}) as $\dot O$ or $\delta A(C) / \delta C(\sigma)$, and perform the partial path integrations, we have the following operator relations:
\bea
i \theta_3 \dot O &=& \left[ O, \oint_C H_1dH_2 \right]=\left[ O, \oint_C dV \right] , \label{Heisenberg eq} \\
\left[X(\sigma, t), A(C) \right] &=&-i \theta_3, \label{CR  for n=3}
\eea
where the vector field $V$ is that introduced by Nambu.  It is also the Clebsch potential in hydrodynamics.
The meaning of the operator relations can be understood from Eq.~(\ref{amplitude}), namely 
\bea
\langle \alpha \mid \hat O_1 \hat O_2 \mid \beta \rangle=\sum_{\gamma}
\langle \alpha \mid \hat O_1 \mid \gamma \rangle \langle \gamma \mid \hat O_2 \mid \beta \rangle \,.
\eea

{}From the commutation relation Eq.~(\ref{CR  for n=3}), we have the following uncertainty relation using the standard method,
\bea
\sqrt{\langle (\Delta X)^2 \rangle} \sqrt{\langle (\Delta A(C))^2 \rangle} \geq \frac{\theta_3}{2},
\eea
where the expectation value means
\bea
\langle \hat O \rangle \propto \sum_{\alpha} \langle \alpha \mid \hat O \mid \alpha \rangle \,.
\eea
This is the uncertainty relation in 3D case and is a generalization of the quantum mechanical uncertainty relation in 2D case in Eq.~(\ref{uncertainty in 2D})

Therefore, the 3D hydrodynamics on the non-commutative space gives the minimum volume of the space equal to $\theta_3/2$, so that the material consisting of the fluid is not a point particle but a particle with a finite volume, or the granular material.
In the general Nambu dynamics with $n$-dimensional phase space, the corresponding uncertainty relation yields
\bea
\sqrt{\langle (\Delta X)^2 \rangle} \sqrt{\langle (\Delta V(C_{n-2}))^2 \rangle} \geq \frac{\theta_n}{2},
\eea
where $V(C_{n-2})$ is the volume of the $(n-2)$-cycle $C_{n-2}$ on which the quantum theory is based.

To make clearer the connection of Nambu dynamics to string (or more extended objects), we will write the action $S_3$ as follows:
\bea
S_3=\int \left[ X(\sigma, t) \frac{\partial (Y, Z)}{\partial (\sigma, t)}- \left(H_1 \frac{\partial}{\partial \sigma} H_2\right) \right] d\sigma dt.
\eea
Then, the Hamiltonian density of the string $\cal H$ reads
\bea
{\cal H}=H_1 \frac{\partial}{\partial \sigma} H_2.
\eea
In the toy model in 3D, 
\bea
{\cal H}= \frac{1}{2} \left(X^2+Y^2+Z^2\right), 
\eea
and so integration over $X$ gives the Lagrangian density of the toy model as
\bea
{\cal L}=\left( \frac{\partial (Y, Z)}{\partial (\sigma, t)}\right)^2- \frac{1}{2} \left(Y^2+Z^2\right).
\eea
If we choose 
\bea
{\cal H}= \frac{1}{2} \left(X^2\right) + \left( \frac{\partial (Z, X)}{\partial (\sigma, t)}\right)^2 +  \left( \frac{\partial (X, Y)}{\partial (\sigma, t)}\right)^2, 
\eea
then we have a string Lagrangian in the Shild gauge, 
\bea
{\cal L}=\left( \frac{\partial (X^{\mu}, X^{\nu})}{\partial (\sigma, t)}\right)^2.
\eea

In the hydrodynamics, however, we have to clarify more explicitly the meaning of Hamiltonian density $\cal H$, or of the Hamiltonian for the string field $\Psi[C; t]$, which is written in terms of the Clebsh potential $V$,
\bea
\hat H= \frac{1}{\theta_3} \int d\sigma dt {\cal H}= \frac{1}{\theta_3} \oint_ C dV .
\eea
For this purpose, the fundamental relations (F1) and (F2), and the superposition of stream functions studied by Nambu in \cite{Nambu} will be important, which moves to incorporate the ensemble averaging and has the affinity with the string field theory as an example.

\section{Conclusions and Discussions}

In this paper the hydrodynamics on non-commutative space has been explored, starting from the formulation of hydrodynamics in terms of the Poisson and Nambu brackets by Y. Nambu \cite{Nambu}.
In particular, in order to introduce the finite size of the space point or the finite size of the element of the fluid, Poisson and Nambu brackets are replaced by the corresponding Moyal brackets.  In this process an parameter $\theta_2$ (dimension of area) or $\theta_3$ (dimension of volume) is introduced in 2D or 3D hydrodynamics, respectively.  They represent the minimum size of area and volume which is acceptable in 2D and 3D spaces.  The hydrodynamics so obtained has an additional term of $O(\theta_{2,3})^2$ which does not exist in the usual Navier-Stokes equation.  
In order to examine whether our hydrodynamics represents the hydrodynamics of the granular materials, we have to compare the computer simulation of our hydrodynamics with the motion of the granular materials.  We will do it in the next work.

To support the replacement of Poisson and Nambu brackets by Moyal brackets, we compare the Moyal product and the expectation value of the operator products in the path integral method. We adopt a toy model in which the most important phase factor, being related to 2D or 3D phase spaces, is kept definitely, but the Hamiltonian is a simple one consisting of the bi-linear terms or the damping factors of the variables.  Moyal products reproduce the path integral expectation values of the toy model.  It is also recognized that the ordering of the Moyal product is related to the certain ordering in the path integral method.  In 2D case, this is the usual time ordering, but in 3D case the ordering is related to the path ordering in $(\sigma, t)$ space.  It is very important to recognize that the Nambu dynamics in 3D is a closed string theory in which temporal development is carried out by the deformation of the closed string $\delta C$.  Moyal product ordering is related to the path ordering along this small closed string $\delta C$ in the path integral method.

To clarify the uncertainty relation when the Nambu dynamics is quantized, we study the path integral quantization.  Using the action of the Nambu dynamics given by Takhatajan \cite{Takhtajan} and Nambu \cite{Nambu H-J}, we demonstrate the 3D case explicitly in terms of the closed string theory.  Then, we can easily read the operator relations from the path integral expression, and clarify the uncertainty relations: 
In 3D case, it is 
\bea
\sqrt{\langle (\Delta X)^2 \rangle} \sqrt{\langle (\Delta A(C))^2 \rangle} \geq \frac{\theta_3}{2},
\eea
where $X$ is a coordinate, and $A(C)$ is the area surrounded by a closed string $C$ depicted on the $(Y, Z)$ plane, being perpendicular to $X$-axis.

In the general Nambu dynamics with $n$-dimensional phase space, the uncertainty relation yields
\bea
\sqrt{\langle (\Delta X)^2 \rangle} \sqrt{\langle (\Delta V(C_{n-2}))^2 \rangle} \geq \frac{\theta_n}{2},
\eea
where $V(C_{n-2})$ is the volume of the $(n-2)$-cycle $C_{n-2}$ on which the quantum theory is based.  

It is very important to examine the various quantization methods of Nambu dynamics, or to examine the quantum analogs of Nambu brackets.  Classical Nambu brackets satisfy a number of relations.  It may be true that depending on the ingredients of the granular materials, different quantization methods should be applied, and also all the relations satisfied by the Nambu brackets may not be required for some materials.  Therefore, it is worthwhile to remind some of the attempts so far done for Nambu brackets.  For this purpose there is a good summary of the studies before 2008.  Please refer to the footnote 2 of the paper by Cheng-Sum Chen et al.~\cite{Chu:2008qv}. 
In Ref.~\cite{Takhtajan}, the Nambu brackets are studied in details and the Moyal product has been also studied.  Modification of the Moyal brackets the so-called Zariski quantization has been observed in finite dimensions~\cite{Dito:1996xr} 
Moyal brackets and the Zariski quantization are a kind of the deformation of the Nambu-Poisson bracket. 
Furthermore, there exists another way of generalizing the matrix commutator~\cite{Curtright:2002fd} in finite dimensions. However, the relation between the algebraic structure and the Bagger-Lambert-Gustavsson (BLG) model~\cite{B-L-G}, which constructs a three-dimensional $\mathcal{N}=8$ superconformal field theory, is not clear at all because the triple commutator cannot meet the fundamental identity. 
In addition, in principle, it is possible to adopt the cubic matrix to describe the 3-algebra~\cite{ALMY-K}, by which, unfortunately, the fundamental identity cannot be satisfied, and is available only for $A_4$ algebra~\cite{Kawamura:2003cw}. 
The Nambu-Poisson bracket with the cut-off representing the Lie 3-algebra in 
finite dimensions proposed in Ref.~\cite{Chu:2008qv} is 
considered to the first attempt meeting the fundamental identity, 
so that it can be compatible with the BLG model. 

After 2008, the M5-brane based on the Nambu-Poisson bracket~\cite{M5-brane} has also been studied. 
Moreover, gauge theories constructed with the Nambu-Poisson bracket have also 
been studied in Ref.~\cite{Jurco:2014aza} 
(for a recent review on the Nambu-Poisson bracket, 
see, e.g.,~\cite{Ho:2013iia}). 
Complete independent basis for structure constants of the volume preserving diffeomorphism (VPD) has been examined~\cite{Sato:2014kva}. 

Finally, we will attempt to rewrite the Nambu dynamics as a matrix model.  Matrix formulation of membrane theory was first carried out by Jens Hoppe in his PhD thesis \cite{Hoppe thesis}.  If the action $S_3$ is invariant under the area-preserving diffeomorphisms in $(\sigma, t)$ space, then his method is applicable.  We combine $\sigma$ and $t$ to $\sigma_a (a=1, 2)$ as $\sigma_1=\sigma$ and $\sigma_2=t$.  Then, the infinitesimal area-preserving transformation reads
\bea
\delta_{\xi}\sigma^a = \{ \sigma^a, \xi(\sigma) \},
\eea
and so it forms an algebra
\bea
\delta_{\xi_1}\delta_{\xi_2}-\delta_{\xi_2}\delta_{\xi_1}=\delta_{\{\xi_1, \xi_2\}}.
\eea
This algebra is shown to be equal to the $N \to \infty$ limit of $SU(N)$ 
in~\cite{Hoppe thesis}, 
so that the $X(\sigma), Y(\sigma)$ and $Z(\sigma)$ as well as $t$ can be replaced by the $N \times N$ hermitian matrices with hat.  Poisson brackets are replaced by the commutator of the corresponding matrices~\cite{BFSS-AIKKTT}, 
\bea
\{A, B\} \to \lim_{N \to \infty} \frac{N}{i} [\hat A, \hat B], 
\eea
and $\int d\sigma dt$ becomes $\left(1/N\right)$ Tr of matrices.
In this way we may arrive at the action of a matrix model,
\bea
S_3= \frac{1}{i} \mbox{Tr} \left( {\hat X} [\hat Y, \hat Z] - {\hat H_1}[{\hat H_2}, {\hat t} \,] \right).
\eea
This expression is, however, far from the correct one, since the area-preserving deffeomorphisms in $(\sigma, t)$ space does not exist or is obscure in the non-relativistic hydrodynamics.  However, being apart from the symmetries in the treatment of $X^{\mu}(\sigma_1, \sigma_2,  \dots, \sigma_D, t)$, if $D=2$, to consider $\sigma_1$ and $\sigma_2$ as indices of row and column is very natural, so that for $D=3$, the appearance of cubic matrix is also natural.  To consider what kind of symmetries may be crucial in studying the hydrodynamics of granular materials, since the symmetry of the ingredients such as of ball, cube or tetrahedron may be partly considered in the symmetry of the variables describing the hydrodynamics.

%
%
\section*{Acknowledgements}
After submission of this paper, the authors were informed relevant papers \cite{Prof-Vancea}, \cite{Prof-Pedraza}, \cite{Curtright:2001jn}, \cite{Curtright:2003je} from
Dr. Ion Vasile Vancea, Dr. Juan F. Pedraza, Dr. Thomas Curtright.
The authors give sincere thanks to them.  This work was partially supported by 
the JSPS Grant-in-Aid for Young Scientists (B) \# 25800136 (K.B.).

%
%

\end{document}